**Assessment of valley coherence in a high-quality monolayer molybdenum diselenide**

Yuto Urano[1,2], Xue Mengsong[1,2], Kenji Watanabe[3], Takashi Taniguchi[1], and Ryo Kitaura[1,*]

[1] *Research Center for Materials Nanoarchitectonics, National Institute for Materials Science, 1-1 Namiki, Tsukuba 305-0044, Japan*

[2] *Department of Chemistry, Nagoya University, Nagoya 464-8601, Japan*

[3] *Research Center for Electronic and Optical Materials, National Institute for Materials Science, 1-1 Namiki, Tsukuba 305-0044, Japan*

Abstract

We investigate the valley coherence in high and low-quality monolayer $MoSe_2$ by polarization-resolved photoluminescence spectroscopy. The observed valley coherence is on the order of 10 % regardless of the sample quality, proving that the suppression of extrinsic effects does not improve the valley coherence. The valley decoherence time estimated based on the valley coherence time and exciton lifetime is sub-picosecond at the longest, which suggests that intrinsic scattering sources, such as phonons, strongly limit the valley coherence.

Quantum bits (qubits) are at the heart of quantum computing [1], which can potentially solve problems that classical computers cannot, such as simulating complex chemical phenomena, breaking encryption codes and solving combinatorial optimization problems. Among the different types of qubits being studied, such as superconducting qubits [2] and trapped-ion qubits [3], semiconductor-based qubits [4] are compatible with existing semiconductor technology, promising large-scale integration of qubits for a general-purpose quantum computing platform. This possibility has motivated research efforts on semiconductor-based qubits [5], including quantum dot qubits [6], silicon-based spin qubits [7] and germanium hole spin qubits [8], over the past decades. Although semiconductor-based quantum computing is still at an early stage, semiconductor-based qubits offer a promising way to realize the potential of quantum computing in the future.

The valley qubit is a semiconductor-based qubit [9] that relies on the valley degree of freedom in certain semiconductors to store and manipulate quantum information. The valley degree of freedom (VDOF) is a property of particular semiconductors that results from two energy extrema (valleys) in the band structure. By encoding quantum information in VDOF, valley qubits offer an alternative platform for quantum computing that complements other qubit platforms, such as spin qubits. For example, in particular types of semiconductors, valley qubits have the potential to be optically manipulated [10], making them an option for future all-optical quantum computing.

Two-dimensional (2D) transition metal dichalcogenides (TMDs) can host valley qubits via excitons with VDOF. [11-13] For example, monolayer $MoS_2$ has two independent K and K' valleys due to inversion symmetry breaking and strong spin-orbit coupling [14]. In particular, right and left circularly polarized light can selectively excite each valley, producing bound electron-hole pairs or excitons with VDOF. [15] Furthermore, the coherent superposition of K and K' excitons can be easily generated when TMDs are excited by linearly polarized light [16-19], the superposition of left and right circularly polarized light. Although this easy optical generation of the coherent K/K' superposition is a great advantage, it has been reported that decoherence of the valley superposition occurs rapidly, even at low temperatures. Therefore, elucidating the factors that cause decoherence is a crucial issue.

Several factors degrade the quality of 2D materials. In particular, environmental factors, including surface adsorbates and substrate effects (roughness, charged impurities, low energy phonons), can lead to additional carrier scattering and significant inhomogeneous broadening at optical transitions. In order to study intrinsic valley coherence in 2D TMDs, it is therefore essential to use a high-quality region free from environmental effects. For this purpose, we have investigated the sample-quality dependence of valley coherence in a 2D TMD, monolayer $MoSe_2$, in this work. We performed polarization-resolved photoluminescence measurements of two regions in the same monolayer $MoSe_2$ flake, a high-quality and a low-quality region. We found that the valley coherence of the two regions is nearly the same, ~ 10 %, although there is a significant difference in the full width at half maximum (FWHM), a measure of sample quality. The upper bound of observed exciton lifetimes in both regions are on the order of picosecond, indicating fast valley decoherence occurs in both cases. Excitation at approximately 1s exciton resonance did not significantly alter valley coherence. These results show that valley decoherence occurs rapidly even when extrinsic scattering is suppressed.

Figure 1(a) shows an optical microscope image of a high-quality MoSe$_2$ monolayer prepared by the graphene-capping-assisted nanosqueezing (GCAN) method [20]; the MoSe$_2$ monolayer is encapsulated by hexagonal boron nitride (hBN) and graphene (G) flakes as G/hBN/MoSe$_2$/hBN/G. The GCAN method relies on AFM tip sweeping to remove impurities and bubbles encapsulated between MoSe$_2$ and hBN. Using the GCAN method, encapsulated impurities and bubbles can be removed without suffering from the random electrostatic potential of charged impurities attached during the squeezing process (Fig.1b). Although there is no difference in the optical image between regions prepared with/without the GCAN method (the orange and dark blue rectangles in Fig. 1(a)), PL spectra show a significant difference. As shown in Fig. 1c, FWHM of the excitonic peak at 1.65 eV of the high-quality region is minimal, 1.7 meV at 10 K (the orange line in Fig. 1c). In contrast, a low-quality region prepared without squeezing shows a much broader peak (the blue line in Fig. 1c); a least-square fitting using single Voigt function gives an FWHM of 6.7 meV. In addition, the PL intensities of the high-quality region are about 5 times stronger than those of the low-quality sample at the same excitation power of 13 µW, demonstrating that non-radiative recombination is strongly suppressed in the high-quality sample [21]. The significant difference in FWHM and peak intensity between the two regions clearly shows that the environmental inhomogeneity due to roughness and impurities is much less in the high-quality region.

We then measured the valley coherence of each region. The regions were excited with linearly polarized light at an excitation energy of 1.72 eV to generate coherent K/K' superpositions, and we performed polarization-resolved PL spectroscopy at 10 K for valley coherence measurements. Figures 2(a) and (b) show polar plots of the polarization direction dependence of the PL intensities of each region, demonstrating a small dependence in the polarization dependence, i.e., a small valley coherence; 90 degrees means that the PL polarization corresponds to the excitation polarization. The valley coherence, $P_c$, can be evaluated as $P_c = (I_+ - I_-)/(I_+ + I_-)$, where $I_+$ and $I_-$ are PL intensities with polarization parallel and vertical to the excitation polarization, respectively. For 1.72 eV excitation, $P_c$ at the squeezed high-quality region in Fig. 1(a) and the other low-quality are 8.8±0.3 % and 11.8±0.9 %, respectively. Regardless of the quality of the sample, the valley coherence is of the order of a few percent, indicating the presence of decoherence paths apart from extrinsic scatterings due to impurities, roughness, etc.; the observed low valley coherence is in agreement with previous reports. [22-24] To exclude the possibility that valley coherence was lost during relaxation to band extremes, we tuned the excitation energy to energies closer to the 1s exciton resonance, 1.68 eV. [25] Figures 3 show polarization-resolved PL spectra of the high and low-quality regions measured with 1.68 eV excitation. Compared to $P_c$ measured with 1.72 eV excitation, both low and high-quality regions show an enhancement of valley coherence by tuning the excitation energy: 8.8±0.3 % (1.72 eV) and 9.4±0.4 % (1.68 eV) at the high-quality region, and 11.8±0.9 % (1.72 eV) and 15.8±0.5 % (1.68 eV) at the low-quality region. Although slight enhancement in $P_c$ was seen at low excitation photon energy, $P_c$ is still on the order of 10 %.

For further insight into the valley coherence of each quality region, we evaluated the valley coherence time, $\tau_{vc}$. First, $\tau_{vc}$ can be related to the valley coherence $P_c$ as $P_c = \tau_{vc} / (\tau_{vc} + \tau)$ [26,27], where $\tau$ is the exciton lifetime. As shown in Fig. 4, the time dependence of the PL intensity measurements has revealed that PL intensities are almost identical to the instrument response function at high and low-quality regions, indicating that both regions show $\tau$ shorter than the time resolution of our measurement system, on the order of $10^0$ ps at the longest; least-square fittings give 1.8 and 7.1 ps for low and high-quality regions, respectively. The observed $P_c$ is typically a few percent, implying that $\tau_{vc}$ is at least an order of magnitude shorter than $\tau$ in both regions. Since extrinsic scattering is minimal in the high-quality region, the observed small $\tau_{vc}$ in the high-quality region should arise from intrinsic scattering sources, such as phonon. [28,29] One way to suppress phonon scattering, in addition to lowering the temperature, is to reduce the dimensionality to zero, i.e., to make quantum dots, where discrete energy levels can suppress phonon scattering, especially scatterings by low-energy phonons.

In summary, we measured the quality dependence of the valley coherence in monolayer $MoSe_2$. The small FWHM observed clearly demonstrates that extrinsic scatterings are strongly suppressed in the high-quality region. However, the valley coherence measured at both high and low-quality regions is on the order of 10 %, indicating that suppression of extrinsic effect does not contribute to enhancing the valley coherence. The valley decoherence time estimated based on the valley coherence and exciton lifetime is sub-picosecond at the longest, regardless of the sample quality. This result suggests that intrinsic source, such as phonon, limits the valley coherence, and reducing dimensionality is probably essential for TMD-based quantum information applications.

**Acknowledgments**
R.K. was supported by JSPS KAKENHI Grant No. JP23H05469, JP22H05458, JP21K18930 and JP20H05664, and JST CREST Grant No. JPMJCR16F3, SCICORP Grant No. JPMJSC2110 and PRESTO Grant No. JPMJPR20A2. K.W. and T.T. acknowledge support from the JSPS KAKENHI (Grant Numbers 20H00354, 21H05233 and 23H02052) and World Premier International Research Center Initiative (WPI), MEXT, Japan.

# Figure captions

Figure1

(a) Optical microscopy images of the graphene/hBN/MoSe$_2$/hBN/graphene sample. The high-quality region and low-quality region are indicated by the orange and dark blue rectangles, respectively. (b) Schematic image showing the nanosqueezing process by AFM tip sweeping. Interlayer bubbles and impurities are removed at the squeezed region. (c) Typical PL spectra taken at the low (blue) and high (orange) quality regions. Both spectra were measured at 10 K with an excitation energy of 1.72 eV.

Figure 2

Polar plots of the polarization direction dependence in the PL intensities of (a) the high-quality and (b) the low-quality regions. In each polar plot, the PL intensity is normalized to the zero polarization direction. Polarization-resolved PL spectra of (c) the high-quality and (d) the low-quality regions. Red and light blue curves correspond to PL spectra with the polarization direction parallel and perpendicular to the incident polarization direction. All PL spectra were measured at an excitation energy of 1.72 eV.

Figure 3

Polarization-resolved PL spectra from (a) the high-quality and (b) the low-quality regions. The red and light blue curves correspond to the PL intensity with the polarization direction parallel and perpendicular to the incident polarization direction. All PL spectra were measured at an excitation energy of 1.68 eV.

Figure 4

Time dependence of PL intensity of low and high-quality regions at 10 K with excitation energy and power of 1.72 eV and 10 μW, respectively. Red and black dots represent the PL intensities of the high and low-quality regions. The red and black lines are fitting curves based on the convolution of the instrument response function and the double exponential decay function.

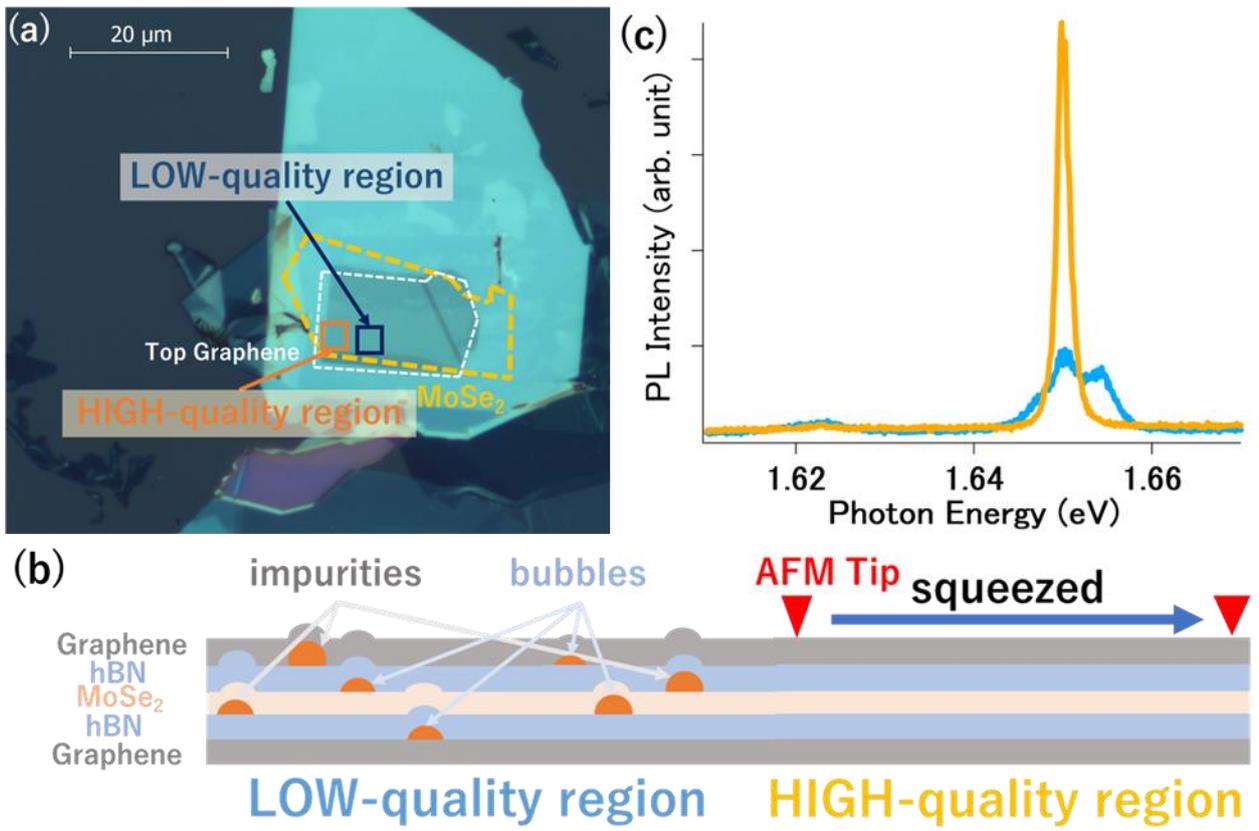

Fig. 1

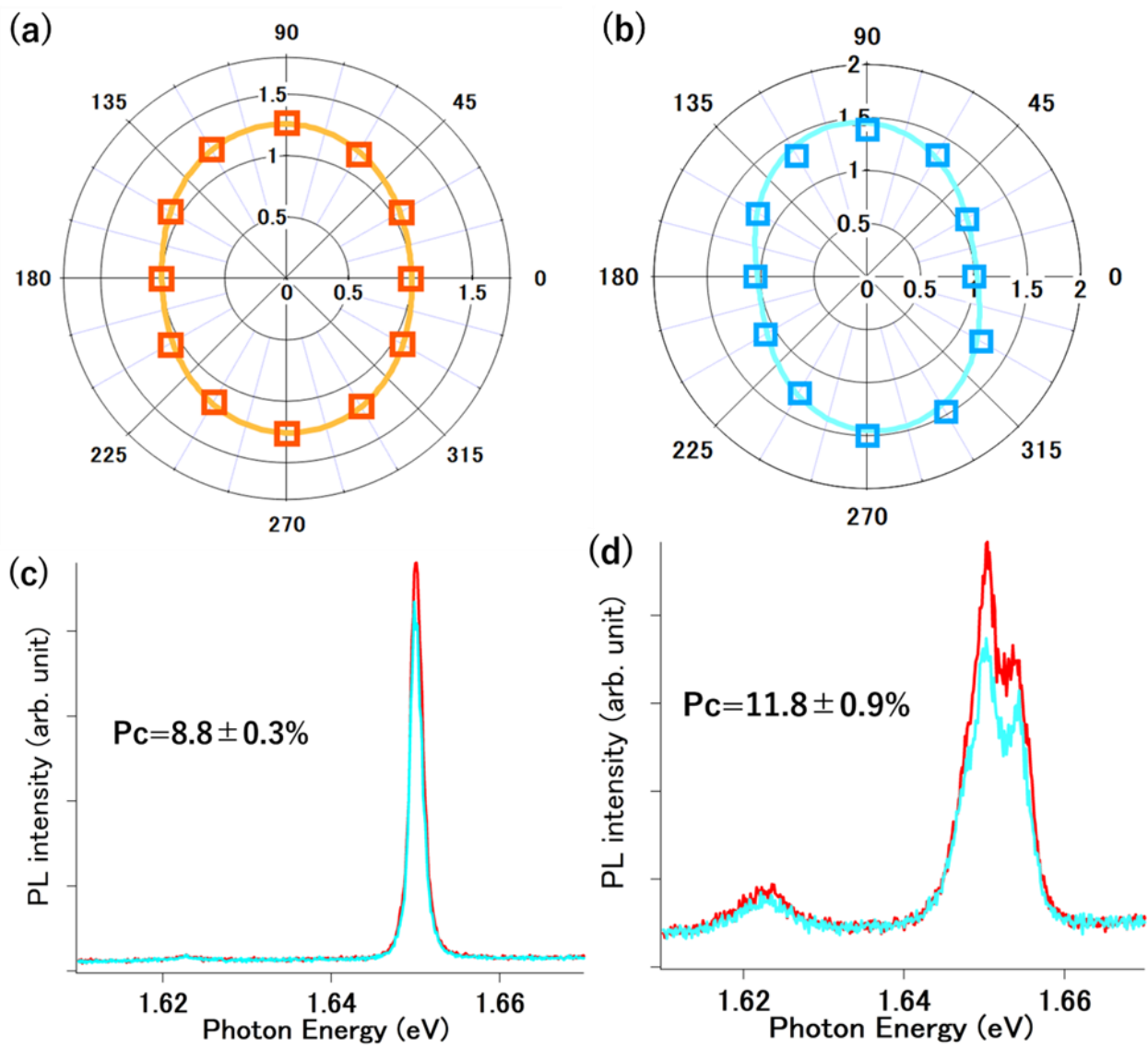

Fig. 2

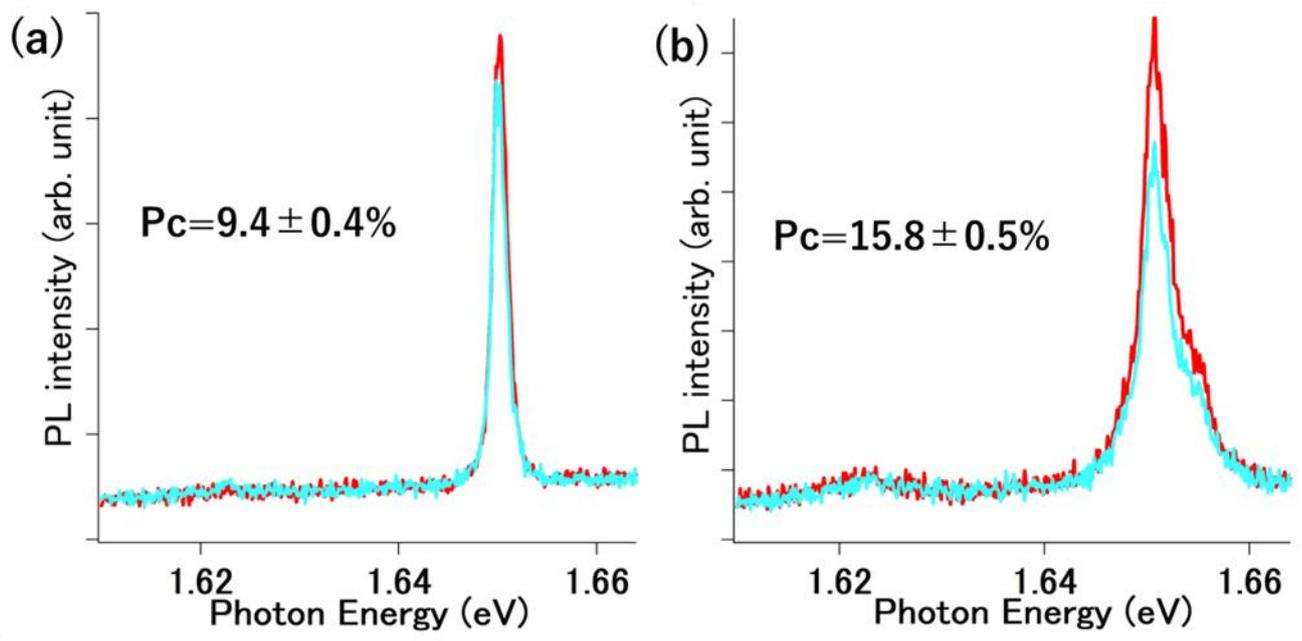

Fig. 3

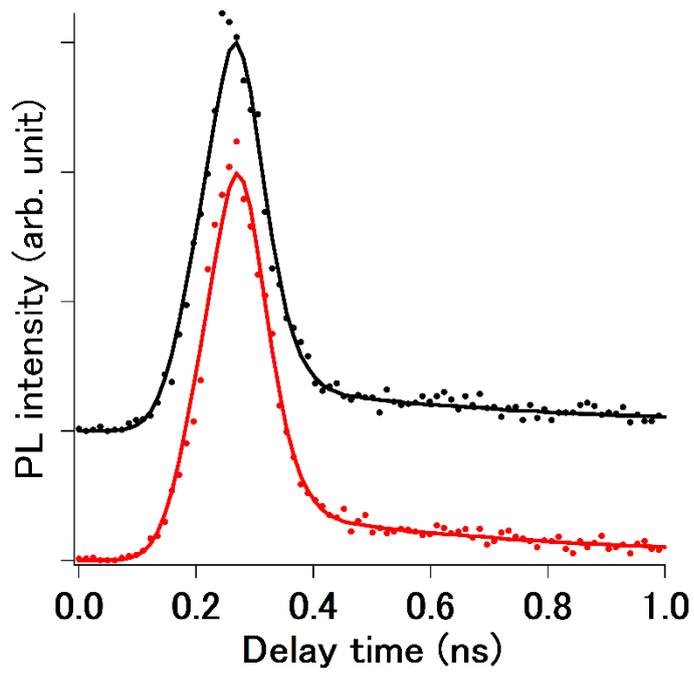

Fig. 4